\documentclass[submission, Proceedings]{SciPost}

\hypersetup{
    colorlinks,
    linkcolor={red!50!black},
    citecolor={blue!50!black},
    urlcolor={blue!80!black}
}

\usepackage[bitstream-charter]{mathdesign}
\DeclareSymbolFont{usualmathcal}{OMS}{cmsy}{m}{n}
\DeclareSymbolFontAlphabet{\mathcal}{usualmathcal}

\begin{document}

\begin{center}{\Large \textbf{
Gluon saturation in proton and its contribution to single inclusive soft gluon production in high energy proton-nucleus collisions\\
}}\end{center}

\begin{center}
Ming Li\textsuperscript{$\star$} 
\end{center}

\begin{center}
Department of Physics, North Carolina State University, Raleigh, NC 27695, USA
\\
*mli48@ncsu.edu
\end{center}

\begin{center}
\today
\end{center}


\definecolor{palegray}{gray}{0.95}
\begin{center}
\colorbox{palegray}{
  \begin{tabular}{rr}
  \begin{minipage}{0.1\textwidth}
    \includegraphics[width=22mm]{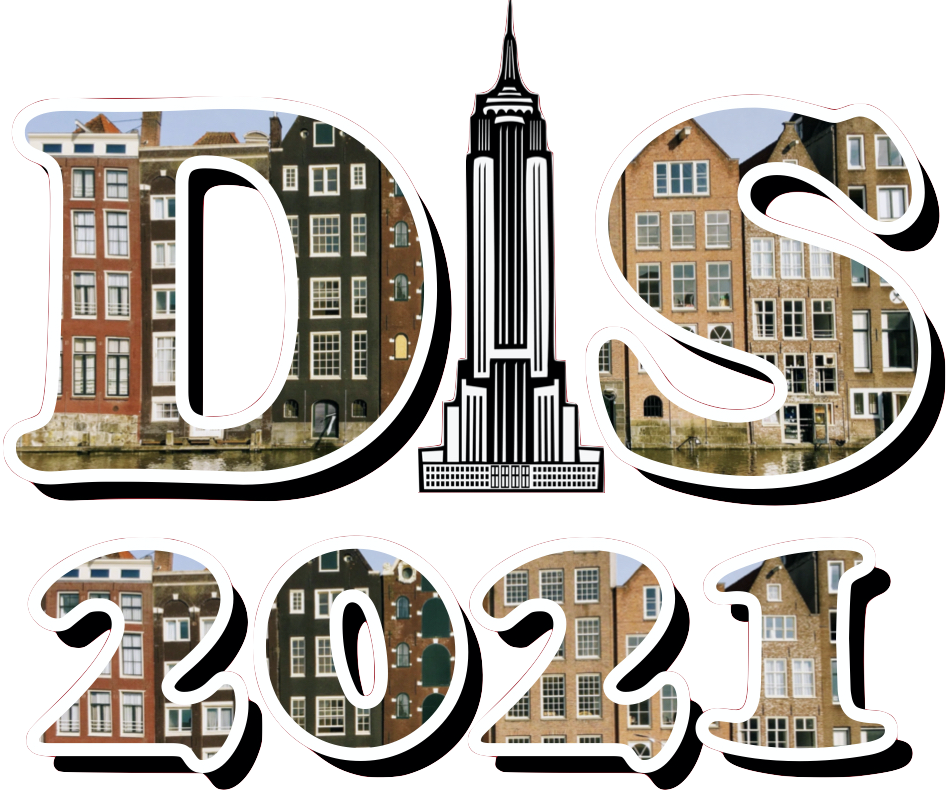}
  \end{minipage}
  &
  \begin{minipage}{0.75\textwidth}
    \begin{center}
    {\it Proceedings for the XXVIII International Workshop\\ on Deep-Inelastic Scattering and
Related Subjects,}\\
    {\it Stony Brook University, New York, USA, 12-16 April 2021} \\
    \doi{10.21468/SciPostPhysProc.?}\\
    \end{center}
  \end{minipage}
\end{tabular}
}
\end{center}

\section*{Abstract}
{\bf
The leading order  single inclusive soft gluon production in high energy proton-nucleus (pA) collisions has been studied by various approaches for more than two decades. The first correction due to the gluon saturation in proton was  analytically attempted recently through a diagrammatic approach in which only partial results were obtained. An important feature of the first saturation correction is that it includes both initial state and final state interactions. In this paper, we present the complete result derived from the Color Glass Condensate framework. Our approach is to analytically solve the classical Yang-Mills equations in the dilute-dense regime and then use the Lehmann-Symanzik-Zimmermann (LSZ) reduction formula to obtain gluon production from classical gluon fields. 
}


\section{Introduction}
\label{sec:intro}
Single inclusive gluon production in  high energy pA collisions plays an important role in understanding the vast amount of experimental data from RHIC and LHC. These include charged particle transverse momentum spectrum as well as multiple particle angular correlation patterns. The leading order result has been studied for more than two decades \cite{Kovchegov:1998bi,Kopeliovich:1998nw,Kovner:2001vi,Dumitru:2001ux}. It treats the proton as a perturbative object while resumming all the multiple interactions with the nucleus eikonally. There are several limitations regarding the leading order result. First, it is symmetric with respect to momentum change $\mathbf{k}\leftrightarrow -\mathbf{k}$ and thus, in contrary to experimental findings, always leads to vanishing triangular flow.  Second, it assumes no final state interaction after scatterings, which might not be a good approximation for describing high multiplicity events.  

\begin{figure}[!t]
    \centering
    \includegraphics[width=0.7\textwidth]{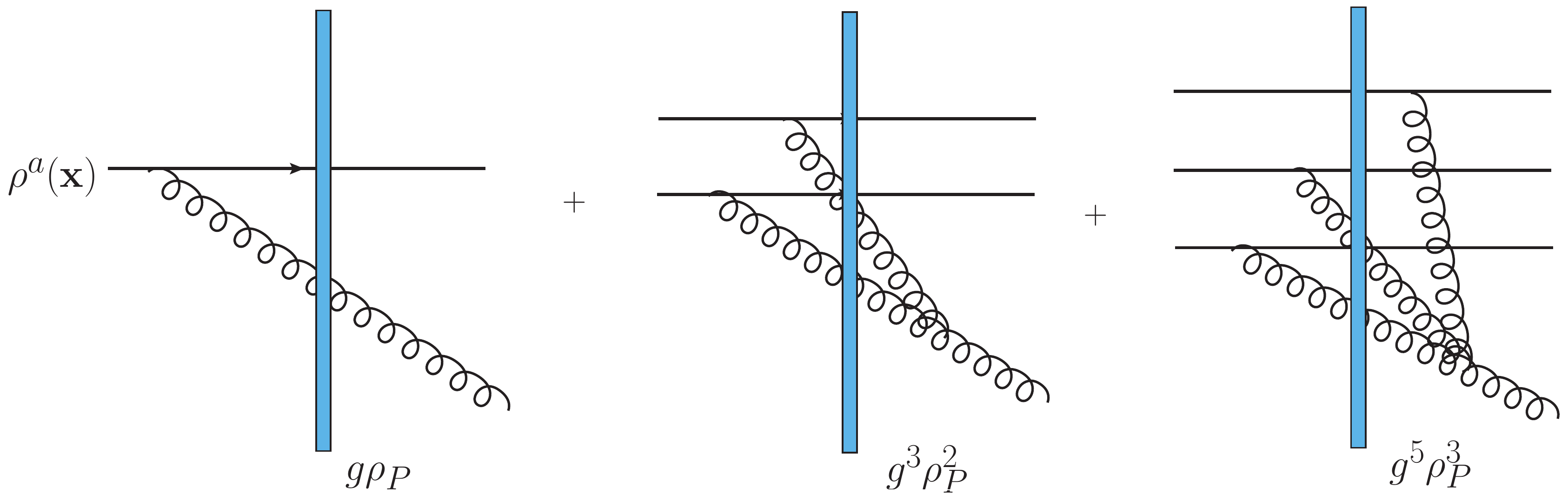}
    \caption{Diagrams illustrating the first saturation correction to single inclusive gluon production. Each horizontal line represents color charge density of proton at different transverse positions. The solid rectanglular bar indicates the Lorentz contracted nucleus. There are tens of diagrams at order $g^3\rho_P^2$ and $g^5\rho_P^3$, only two representative diagrams are shown here.  } 
	\label{fig:f1}
\end{figure}

Motivated by these considerations, next to leading order corrections, specifically corrections due to gluon saturation effect in proton are studied. The saturation correction is different from general perturbative corrections which are usually  organized in powers of coupling constant $g$. Figure \ref{fig:f1} shows some representative diagrams corresponding to the first saturation correction which, at fixed order of $g$, only capture terms that are enhanced by the color charge density of the proton $\rho_P^a(\mathbf{x})$. For example, at order $g^3$ we only consider diagrams enhanced by $g^3\rho_P^2$ and discard diagrams merely enhanced by $g^3\rho_P$. 

Formally one can write the single gluon production amplitude as 
\begin{equation}
\mathcal{M} = \mathcal{M}_{(1)} + \mathcal{M}_{(3)}+\mathcal{M}_{(5)}+ \ldots
\end{equation}
Here $\mathcal{M}_{(1)}$ corresponds to diagrams at order $g\rho_P$ while $\mathcal{M}_{(3)}$ and $\mathcal{M}_{(5)}$ representing diagrams at order $g^3\rho_P^2$ and $g^5\rho_P^3$, respectively. Both $\mathcal{M}_{(3)}$ and $\mathcal{M}_{(5)}$ contribute to the first saturation correction. Previous studies wre only able to compute $\mathcal{M}_{(3)}$ \cite{Balitsky:2004rr, Chirilli:2015tea, McLerran:2016snu}. For the first time, we have obtained $\mathcal{M}_{(5)}$ \cite{Li:2021zmf,Li:2021yiv}. The amplitude squared is 
\begin{equation}
|\mathcal{M}|^2 =|\mathcal{M}_{(1)}|^2 + \mathcal{M}_{(1)}^{\ast} \mathcal{M}_{(3)} + \mathcal{M}_{(1)}\mathcal{M}_{(3)} ^{\ast} + |\mathcal{M}_{(3)}|^2 + \mathcal{M}_{(1)}^{\ast} \mathcal{M}_{(5)} + \mathcal{M}_{(1)}\mathcal{M}_{(5)} ^{\ast}+ \ldots
\end{equation}
The leading order result comes from $|\mathcal{M}_{(1)}|^2$. The next two terms $\mathcal{M}_{(1)}^{\ast} \mathcal{M}_{(3)}+c.c$ vanish upon ensemble averaging over the color charge density configurations of the proton using Gaussian like models such as the McLerran-Venugopalan model. It does not contribute to single gluon production but will contribute to double and multiple gluon productions. The first saturation correction therefore is 
\begin{equation}\label{eq:fsc_formula}
\frac{dN}{d^2\mathbf{k}}\Big|_{FSC} = |\mathcal{M}_{(3)}|^2 + \mathcal{M}_{(1)}^{\ast} \mathcal{M}_{(5)} + \mathcal{M}_{(1)}\mathcal{M}_{(5)} ^{\ast}
\end{equation}
It is proportional to $g^6\rho_P^4$ as compared to leading order result $g^2\rho_P^2$.

How do we compute the single gluon production amplitude? We work in the Color Glass Condensate (CGC) framework. First, we obtain the classical gluon fields produced in high energy pA collisions by solving the classical Yang-Mills equations in the dilute-dense regime.  Second, using the LSZ reduction formula, we define the asypmtotic gluon creation operators for the two independent polarization modes as 
\begin{equation}
\begin{split}
\hat{a}_{\eta}^{a \dagger}(\mathbf{k})
=& -i\tau \sqrt{\frac{\pi}{4}}\left( H_1^{(2)}(k_{\perp}\tau) \overleftrightarrow{\partial_{\tau}} \tilde{\beta}^a(\tau, \mathbf{k})\right)\Big|_{\tau=+\infty},\\
\hat{a}_{\perp}^{a\dagger}(\mathbf{k}) =& - i\tau \sqrt{\frac{\pi}{4}}\left( H_0^{(2)}(k_{\perp}\tau) \overleftrightarrow{\partial_{\tau}} \beta^a_{\perp}(\tau, \mathbf{k})\right)\Big|_{\tau =+\infty}.\\
\end{split}
\end{equation}
Here $\tilde{\beta}^a(\tau, \mathbf{p})$ and $\beta_{\perp}^a(\tau, \mathbf{p})$ are the two independent scalar modes of classical gluon fields in momentum space. Their expressions will be given in the next section.  The $H^{(2)}_0(x)$ and $H^{(2)}_1(x)$ are Hankel  functions. The derivative is defined as  $f_1(\tau)\overleftrightarrow{\partial_{\tau}} f_2(\tau) = f_1(\tau) \partial_{\tau} f_2(\tau) - \partial_{\tau} f_1(\tau) f_2(\tau)$. The single gluon production thus can be computed by
\begin{equation}
\frac{dN}{d^2\mathbf{k}} = \frac{1}{(2\pi)^2} \Big(\hat{a}_{\eta}^{a \dagger}(\mathbf{k})\hat{a}_{\eta}^{a}(\mathbf{k})+ \hat{a}_{\perp}^{a \dagger}(\mathbf{k})\hat{a}_{\perp}^{a}(\mathbf{k}) \Big).
\end{equation}
To compute the first saturation correction eq. \eqref{eq:fsc_formula}, it turns out we only need the next to leading order solutions of classical Yang-Mills equations $\tilde{\beta}_{(3)}(\tau, \mathbf{x})$ and $\beta_{\perp (3)}(\tau, \mathbf{x})$.

\section{Classical Gluon Fields at Next to Leading Order}
To obtain the classical gluon fields produced in high energy pA collisions, one solves the classical Yang-Mills equations in the forward light cone. We work in the Fock-Schwinger gauge $x^-A^+ + x^+A^-=0$ so that one can parameterize the solutions as $A^+ = x^+ \beta$, $A^- =-x^- \beta$ and $A^i=\beta^i$. We also assume boost invariance and the solutions are independent of spatial rapidity $\eta$. The Yang-Mills equations 
\begin{equation}
D_{\mu} F^{\mu\nu} =0
\end{equation}
are supplemented with the initial conditions 
\begin{equation}
\beta(\tau = 0, \mathbf{x}) = \frac{ig}{2} \left[ \beta^i_P(\mathbf{x}), \beta^i_T(\mathbf{x})\right], \qquad \beta^i(\tau=0, \mathbf{x}) = \beta^i_P(\mathbf{x}) + \beta^i_T(\mathbf{x}).
\end{equation}
Here $\beta^i_P(\mathbf{x})$ and  $\beta^i_T(\mathbf{x})$ are the Weizsacker-Williams gluon fields of the proton and the nucleus before the collisions, respectively. In the dilute-dense regime, we treat the proton as perturbative and solve the equations perturbatively 
\begin{equation}
\beta(\tau,\mathbf{x}) = \sum_{n=0}^{\infty} g^n \beta^{(n)}(\tau, \mathbf{x}),\qquad \beta_i(\tau, \mathbf{x}) =\sum_{n=0}^{\infty} g^n \beta_i^{(n)}(\tau, \mathbf{x}).
\end{equation}
Note that both the equations and the initial conditions are to be expanded and solved order by order. One will need the method of variation of parameters to solve inhomogeneous Bessel equations. Furthermore, a critical mathematical trick needed is Graf's formula that expresses a product of two Bessel functions in terms of angular integral of one Bessel function.  The final next to leading order solutions are  
\begin{equation}\label{eq:g3_solutions_1}
\begin{split}
\beta^{(3)}(\tau,\mathbf{k}) 
=&2\beta^{(3)}(\tau=0,\mathbf{k}) \frac{J_1(k_{\perp}\tau) }{k_{\perp}\tau}-i \int \frac{d^2\mathbf{p}}{(2\pi)} \Big[b_{\perp}(\mathbf{p}), b_{\eta}(\mathbf{k}-\mathbf{p})\Big]\frac{\mathbf{k}\times \mathbf{p}}{p_{\perp}^2|\mathbf{k}-\mathbf{p}|^2}\\
&\qquad \times \int_{-\pi}^{\pi}\frac{d\phi}{2\pi}\Big(1+\frac{2\mathbf{k}\cdot(\mathbf{k}-\mathbf{p})}{w_{\perp}^2-k_{\perp}^2}\Big)\left(\frac{J_1(w_{\perp}\tau)}{w_{\perp}\tau}-\frac{J_1(k_{\perp}\tau)}{k_{\perp}\tau}\right),\\
\end{split}
\end{equation}
\begin{equation}\label{eq:g3_solutions_2}
\begin{split}
\beta^{(3)}_{\perp}(\tau, \mathbf{k}) = &\beta_{\perp}^{(3)}(\tau=0, \mathbf{k})J_0(k_{\perp}\tau) +\frac{i}{k_{\perp}} \int \frac{d^2\mathbf{p}}{(2\pi)^2}\Big[b_{\eta}(\mathbf{p}), b_{\eta}( \mathbf{k}-\mathbf{p})\Big] \frac{ \mathbf{k}\times \mathbf{p}}{2p^2_{\perp}|\mathbf{k}-\mathbf{p}|^2}\\
&\times\int_{-\pi}^{\pi} \frac{d\phi}{2\pi}\Big(1 + \frac{2\mathbf{p}\cdot(\mathbf{k}-\mathbf{p})}{w_{\perp}^2-k_{\perp}^2}\Big)(J_0(w_{\perp}\tau)-J_0(k_{\perp}\tau))-\frac{i}{k_{\perp}} \int \frac{d^2\mathbf{p}}{(2\pi)^2} \Big[b_{\perp}(\mathbf{p}), b_{\perp}(\mathbf{k}-\mathbf{p})\Big]\\
&\times \frac{ (\mathbf{k}\times \mathbf{p})(-\mathbf{p}\cdot\mathbf{k}+p_{\perp}^2+k_{\perp}^2)}{p^2_{\perp}|\mathbf{k}-\mathbf{p}|^2} \int_{-\pi}^{\pi} \frac{d\phi}{2\pi}\frac{1}{w_{\perp}^2-k_{\perp}^2} \left(J_0(w_{\perp}\tau)-J_0(k_{\perp}\tau)\right),\\
\end{split}
\end{equation}
\begin{equation}\label{eq:g3_solutions_3}
\begin{split}
\Lambda^{(3)}(\tau, \mathbf{k}) 
=& - \frac{i}{k_{\perp}^2} \int\frac{d^2\mathbf{p}}{(2\pi)^2}\Big[b_{\eta}(\mathbf{p}), b_{\eta}(\mathbf{k}-\mathbf{p})\Big]  \frac{\mathbf{k}\cdot(\mathbf{k}-2\mathbf{p})}{4p_{\perp}^2 |\mathbf{k}-\mathbf{p}|^2} \int_{-\pi}^{\pi}\frac{d\phi}{2\pi}\Big(1-\frac{p_{\perp}^2 + |\mathbf{k}-\mathbf{p}|^2}{w^2_{\perp}} \Big)(1- J_0(w_{\perp}\tau))\\
& - \frac{i}{k_{\perp}^2} \int\frac{d^2\mathbf{p}}{(2\pi)^2} \Big[b_{\perp}(\mathbf{p}), b_{\perp}(\mathbf{k}-\mathbf{p})\Big] \frac{\mathbf{k}\cdot(\mathbf{k}-2\mathbf{p})\mathbf{p}\cdot(\mathbf{k}-\mathbf{p})}{2p_{\perp}^2 |\mathbf{k}-\mathbf{p}|^2} \int_{-\pi}^{\pi}\frac{d\phi}{2\pi}\frac{1}{w^2_{\perp}}(1- J_0(w_{\perp}\tau)).\\
\end{split}
\end{equation}
We have made the decomposition $\beta_i(\tau, \mathbf{k}) = \frac{-i\epsilon^{ij}\mathbf{k}_j}{k_{\perp}} \beta_{\perp}(\tau, \mathbf{k}) -i\mathbf{k}_i \Lambda(\tau, \mathbf{k})$ in which $\Lambda(\tau, \mathbf{k})$ is a non-dynamical field. The $b_{\perp}(\mathbf{p}) = -i\epsilon_{ij}\mathbf{p}_i\beta_j^{(1)}(\tau=0, \mathbf{p})$ and $b_{\eta}(\mathbf{p}) = 2\beta^{(1)}(\tau=0, \mathbf{p})$ are related to the leading order initial condition. In addition, here $w_{\perp} = \sqrt{p_{\perp}^2+|\mathbf{k}-\mathbf{p}|^2 -2p_{\perp}|\mathbf{k}-\mathbf{p}|\cos\phi}$. 
In the above solutions, terms containing commutators represent final state interactions due to three gluon vertices. Initial state interactions are included in higher order initial conditions $\beta^{(3)}(\tau=0, \mathbf{k})$ and $\beta_{\perp}^{(3)}(\tau=0, \mathbf{k})$. Another important property of the solutions is that the color structure and the time dependence are factorized. These solutions can also be used to compute other  physical quantities of interest such as energy-momentum tensor and angular-momentum tensor.

\section{Results: First Saturation Correction to Single Gluon Production}
Using the next to leading order solutions for gluon fields eqs. \eqref{eq:g3_solutions_1}, \eqref{eq:g3_solutions_2}, one can apply the LSZ reduction formula to compute the first saturation correction to single gluon production, the final results are 
\begin{equation}\label{eq:M3M3}
\begin{split}
&|\mathcal{M}_{(3)}(\mathbf{k})|^2\\
=&-\frac{1}{\pi}\int_{\mathbf{p},\mathbf{p}_1,\mathbf{q},\mathbf{q}_1} \mathcal{H}_1(\mathbf{p},\mathbf{p}_1,\mathbf{q},\mathbf{q}_1,\mathbf{k})\rho_P^{b_1}(\mathbf{p}-\mathbf{p}_1)T^b_{b_1b_2}\rho_P^{b_2}(\mathbf{p}_1)\rho_P^{d_1}(\mathbf{q}-\mathbf{q}_1)T^d_{d_1d_2}\rho_P^{d_2}(\mathbf{q}_1)\\
&\qquad \qquad \times \Big[U(\mathbf{k}-\mathbf{p})U^{T}(-\mathbf{k}-\mathbf{q})\Big]^{bd}\\
&-\frac{1}{\pi}\int_{\mathbf{p},\mathbf{p}_1,\mathbf{p}_2, \mathbf{q},\mathbf{q}_1}\mathcal{H}_2(\mathbf{p},\mathbf{p}_1,\mathbf{p}_2, \mathbf{q},\mathbf{q}_1, \mathbf{k})  \rho_{P}^{b_1}(\mathbf{p}_1)\rho_P^{b_2}(\mathbf{p}_2) \rho_P^{d_1}(\mathbf{q}-\mathbf{q}_1)T^d_{d_1d_2}\rho_P^{d_2}(\mathbf{q}_1)\\
&\qquad\qquad  \times  \Big[U(\mathbf{k}-\mathbf{p}-\mathbf{p}_1)T^aU^{T}(\mathbf{p}-\mathbf{p}_2)\Big]^{b_1b_2}U^{da}(-\mathbf{k}-\mathbf{q})+c.c.\\
&-\frac{1}{\pi}\int_{\mathbf{p},\mathbf{p}_1,\mathbf{p}_2, \mathbf{q},\mathbf{q}_1,\mathbf{q}_2}\mathcal{H}_3(\mathbf{p},\mathbf{p}_1, \mathbf{p}_2,\mathbf{q},  \mathbf{q}_1, \mathbf{q}_2, \mathbf{k})   \rho_P^{b_1}(\mathbf{p}_1)\rho_P^{b_2}(\mathbf{p}_2)\rho_P^{d_1}(\mathbf{q}_1)  \rho_P^{d_2}(\mathbf{q}_2)\\
&\qquad\qquad \times \Big[U(\mathbf{k}-\mathbf{p}-\mathbf{p}_1)T^aU^{T}(\mathbf{p}-\mathbf{p}_2)]^{b_1b_2} \Big[U(-\mathbf{k}-\mathbf{q}-\mathbf{q}_1) T^aU^T(\mathbf{q}-\mathbf{q}_2)\Big]^{d_1d_2}.\\
\end{split}
\end{equation}
and 
\begin{equation}\label{eq:M1M5}
\begin{split}
&\mathcal{M}_{(1)}^{\ast} (\mathbf{k})\mathcal{M}_{(5)}(\mathbf{k}) + \mathcal{M}_{(1)}(\mathbf{k})\mathcal{M}_{(5)} ^{\ast} (\mathbf{k})\\
=&-\frac{1}{\pi}\int_{\mathbf{p},\mathbf{q},\mathbf{p}_1,\mathbf{p}_4}\mathcal{F}_1(\mathbf{p}, \mathbf{q}, \mathbf{p}_1, \mathbf{p}_4, \mathbf{k}) \rho_P^{b_1}(\mathbf{p}_1)T^a_{b_1b_2}\rho_P^{b_2}(\mathbf{p}-\mathbf{p}_1) \rho_P^{b_3}(\mathbf{q}-\mathbf{p})\rho_P^{b_4}(\mathbf{p}_4)\\
&\qquad\qquad \times \Big[T^aU(\mathbf{k}-\mathbf{q})U^T(-\mathbf{k}-\mathbf{p}_4)\Big]^{b_3b_4}\\
&-\frac{1}{\pi}\int_{\mathbf{p},\mathbf{q},\mathbf{p}_1,\mathbf{p}_3,\mathbf{p}_4}\mathcal{F}_2(\mathbf{p}, \mathbf{q}, \mathbf{p}_1, \mathbf{p}_3, \mathbf{p}_4, \mathbf{k}) \rho_P^{b_1}(\mathbf{p}-\mathbf{p_1})T^b_{b_1b_2}\rho_P^{b_2}(\mathbf{p}_1)\rho_P^{b_3}(\mathbf{p}_3)\rho_P^{b_4}(\mathbf{p}_4)\\
&\qquad\qquad \times U^{ba}(\mathbf{q}-\mathbf{p}) \Big[U(\mathbf{k}-\mathbf{q}-\mathbf{p}_3)T^aU^{T}(-\mathbf{k}-\mathbf{p}_4)\Big]^{b_3b_4}\\
&-\frac{1}{\pi}\int_{\mathbf{q},\mathbf{p},\mathbf{p}_1,\mathbf{p}_2,\mathbf{p}_3,\mathbf{p}_4}\mathcal{F}_3(\mathbf{p}, \mathbf{q}, \mathbf{p}_1, \mathbf{p}_2, \mathbf{p}_3, \mathbf{p}_4, \mathbf{k})\rho^{b_1}_P(\mathbf{p}_1)\rho^{b_2}_P(\mathbf{p}_2)\rho^{b_3}_P(\mathbf{p}_3)\rho^{b_4}_P(\mathbf{p}_4)\\
&\qquad\qquad\times\Big[ U(\mathbf{p}-\mathbf{p}_1)T^aU^{T}(\mathbf{q}-\mathbf{p}-\mathbf{p}_2)\Big]^{b_1b_2} \Big[U(\mathbf{k}-\mathbf{q}-\mathbf{p}_3) T^aU^{T}(-\mathbf{k}-\mathbf{p}_4)\Big]^{b_3b_4}\\
&+c.c.
\end{split}
\end{equation}
They are expressed as functionals of the proton color charge density $\rho_P^a(\mathbf{p})$ and nucleus color charge density $\rho_T^a(\mathbf{p})$ (through the adjoint Wilson line $U^{cd}(\mathbf{p})$). The explicit expressions for the kinematic factors $\mathcal{H}_{1,2,3}$ and $\mathcal{F}_{1,2,3}$ are given in \cite{Li:2021yiv}. They are functions of transverse momenta independent of $\rho_P^a(\mathbf{p})$ and $U^{cd}(\mathbf{p})$.  We also used the shorthand notation $\int_{\mathbf{p}} = \int \frac{d^2\mathbf{p}}{(2\pi)^2}$. 

\section{Conclusion}
We have obtained the first saturation correction  to single inclusive soft gluon production in high energy pA collisions. It incorporates both initial state interactions and final state interactions within the proton.  The functional form eqs. \eqref{eq:M3M3} and \eqref{eq:M1M5} in terms of color charge densities could be directly used to compute double and multiple gluon productions. Further evaluations of the saturation correction  require ensemble averaging over products of Wilson lines. These can be done either under some appropriate approximations (such as dipole approximation, large $N_c$ approximation, glasma graph approximation) or through numerical simulations.  On the other hand, further analysis of the the first saturation correction might provide insights on how to compute and resum higher order saturation corrections, which is ultimately related to the unsolved problem of computing single gluon production in high energy nucleus-nucleus collisions.

\section*{Acknowledgements}
I thank V. Skokov for  collaborating on this project and  A. Kovner, Y. Kovchegov, M. Lublinsky, H. Duan for insightful discussions.
\paragraph{Funding information}
We acknowledge support by the DOE Office of Nuclear Physics through Grant No. DE-SC0020081
\bibliography{saturation.bib}
\nolinenumbers

\end{document}